\documentclass[a4paper,10pt]{article}

\newif\ifpdf            
\ifx\pdfoutput\undefined \pdffalse \else \pdftrue \fi
\ifpdf 
\usepackage[pdftex]{graphicx}
\pdfcompresslevel9 \DeclareGraphicsExtensions{.jpg,.pdf,.png,.mps}
\else 
\usepackage[dvips]{graphicx}
\DeclareGraphicsExtensions{.eps,.ps} \fi

\usepackage{graphics}
\usepackage{bm}
\usepackage{amsmath}

\def\mat#1{\ensuremath{\underline{\underline{{\bm{#1}}}}}}

\begin{document}

\title{Networks of helix-forming polymers}

\author{Samuel Kutter and Eugene M. Terentjev\\
Cavendish Laboratory, University of Cambridge,\\
Madingley Road, Cambridge CB3 0HE, UK}

\date{September 2002}
\maketitle

\begin{abstract}\noindent
Biological molecules can form hydrogen bonds between nearby residues,
leading to helical secondary structures. The associated reduction of
configurational
entropy leads to a temperature dependence of this effect: the {\em helix-coil
transition}. Since the formation of helices implies a dramatic shortening of
the polymer dimensions, an externally
imposed end-to-end distance $R$ affects the equilibrium helical fraction
of the polymer and
the resulting force-extension curves show anomalous  plateau regimes.
In this article, we investigate the behaviour of a cross\-link\-ed
network of such helicogenic molecules, particularly
focusing on the coupling of the (average) helical content present in a
network to the externally imposed strain. We show that both elongation and
compression can lead to an increase in helical domains under appropriate
conditions.
\end{abstract}

\vspace{.2cm}
\noindent{PACS numbers:}\\
{78.20.Ek}\ {Optical properties of bulk materials and thin films:
optical activity}\\
{83.80.Va}{Elastomeric polymers}\\
{87.15.La}\ {Biomolecules: structure and physical properties:
mechanical properties}

\section{Introduction}

Recent advances in experimental techniques allow to perform single mo\-le\-cule
experiments on polymer chains. Typical experiments, for example with an atomic
force microscope
\cite{smith-finzi92,rief-oesterhelt97,oesterhelt-rief99,rief-grubmuller02},
reveal a characteristic
force-extension curve of a mo\-le\-cu\-le. However, theoretical understanding is
often limited, if not poor, due to  the complicated nature of interactions
between individual parts of the mo\-le\-cu\-le: the force-extension behaviour
strongly depends on the microscopic configurations of the polymer, which, for
complex biological mo\-le\-cu\-les can exhibit a very rich energy landscape
with respect to configurational coordinates. In this case, one expects a strong
dependence on the actual pathways of unfolding and folding respectively.
Furthermore, the experimental manipulation might influence the dynamics of the
mo\-le\-cule in this energy landscape such that the ordinary equilibrium
statistical mechanics breaks down and the system behaves non-ergodically.

\begin{figure} \center
\resizebox{0.6\textwidth}{!}{\includegraphics{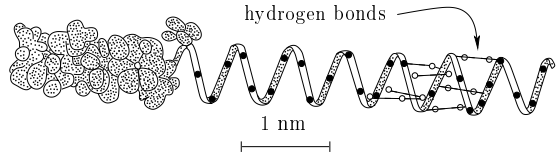}} \caption{An
$\alpha$-helix of a polymer, showing in a schematic way the van der
Waals radii on the left, the backbone in the center and the hydrogen bonds
between residues on the right, after \cite{grosberg-khokhlov94}.} \label{fig1}
\end{figure}

However, certain ho\-mo\-po\-ly\-pep\-ti\-des form long strands of helical
segments of regular $\alpha$-helices under
appropriate conditions (see fig. \ref{fig1}), hence the folding pathways are
very simple in this case: on a coarse grained level, the
internal states of these mo\-le\-cu\-les are described according to the
Zimm-Bragg model \cite{zimm-bragg59}: it assumes that each segment along the
polymer backbone has access to merely two states, a random coil-like, unbound
state and a helical state, where the particular residue forms a hydrogen bond
with specific other residues at a certain distance along the backbone. The
state of the
polymer can therefore be described by a simple sequence $\{hhcchchc\dots\}$,
where $c$ and $h$ stand for the coil or helical state of each listed segment.
The interactions come into play by the use of an appropriate Hamiltonian, which
can be of arbitrary sophistication, for example taking into account the fact
that segments can only form a H-bond with, for instance, their fourth neighbours
\cite{poland-scheraga70,saroff-kiefer99,bloomfield99}. This allows to compute
key properties of a molecule, e.g. the average helical fraction in a chain, by
applying standard me\-thods of statistical mechanics.
Experimentally, one can determine the fraction of monomers in a helical state by
measuring the optical activity in a solution of helix-forming chains.
Since, due to the coherent ordering,
helical domains rotate the polarisation of light much more 
strongly than the individual chiral monomers in the coil state,
the measure of optical activity gives a direct indication of the fraction of
monomers in the helical state \cite{teramoto01,yue-berry96,doty-bradbury56}.

This approach can be extended to deal with the case
where the end-to-end distance
$R$ is constrained externally. Recently, two different groups
\cite{buhot-halperin02,buhot-halperin00,tamashiro-pincus01} have independently
calculated the effects of such a
constraint on the force and the helical fraction, revealing large plateaus in
the force-extension curve and the formation of helical domains upon
stretching. In contrast to the end-to-end distance, which can be controlled
experimentally, the helical fraction cannot be determined for a single molecule.
On the other hand, as mentioned above, the average helical fraction of solutions
is perfectly accessible, whereas now the end-to-end distance of individual
molecules
is left uncontrolled. In order to determine the relationship between the
helical fraction and the end-to-end distance, one needs to be able to control
the end-to-end distance and to simultaneously measure the helical content.

We show that a network made of helix-forming polymer
mo\-le\-cu\-les can provide a system which fulfills
both requirements: a macroscopic deformation is experimentally easy to impose
and affects the end-to-end distance of network strands. At the same
time, the optical activity can be monitored, provided the sample is sufficiently
transparent.

However, there is a price to pay: random polymer networks are very
complex and poorly understood systems: topological constraints alter the
statistics due to the fact that the end-points of strands are cross\-link\-ed
to each other and different chain segments cannot intersect.
The cross\-link\-ed
network remembers its history: its properties depend on the conditions
at the moment of its formation.

Although any network theory faces numerous difficulties, the most simple
approach of the {\em phantom chain network} is frequently and successfully
used. In
this model, the chains are assumed to interact at their endpoints only, i.e.
at the crosslinks. The model neglects the interaction along the backbone, and
therefore assumes the chains to be transparent to each other and free
to intersect. Despite its simplicity, the model reproduces many aspects of the
experimental data \cite{treloar75}.

In this article, we unite two approaches, the
statistical mechanics of a single helix-forming mo\-le\-cu\-le
and a basic network theory. We base our investigations on the ideas by
Buhot and Halperin \cite{buhot-halperin02,buhot-halperin00}
for the response of a single helix-forming
molecule to an externally imposed end-to-end extension and on the phantom chain
network approach to therefrom calculate the network properties.

Our approach not only provides a useful
extension of the single chain theory,
but also  offers a way to control the helical content in the network. By
deforming rubber or gel, one indirectly affects the end-to-end distances of
the polymer strands. As they are stretched or compressed, the average
helical fraction changes. Hence, in an optically transparent gel network,
one can determine the helical
fraction by  measuring the optical activity (as in the case of a solution
\cite{teramoto01,yue-berry96,doty-bradbury56}),
and moreover by imposing a macroscopic deformation
one can identify the coupling between mechanical and optical
responses, and thus draw conclusions about the effect of deformation on the
helical fraction.

The article is organised as follows: in section \ref{sec2}, we review the
classical Zimm-Bragg model \cite{zimm-bragg59} and the solution me\-thod
proposed by Buhot and Halperin \cite{buhot-halperin02}. In section \ref{sec3},
we briefly outline the {\em phantom chain network}\/ approach to calculate
network properties on the basis of single mo\-le\-cu\-le parameters. Section
\ref{sec-netw} links the approaches of section \ref{sec2} and \ref{sec3}
together and provides a network theory of helix-forming mo\-le\-cu\-les.
Analytical and numerical results, and their discussion follow in section
\ref{sec-r-d}.

\section{Helix-coil transition models}\label{sec2}

Our aim is to describe a network of polymers which can fold into regular
$\alpha$-helices,
due to the formation of regularly spaced hydrogen bonds. This section
reviews the principles of
the helix-coil transition, which are relevant to our investigations. In the
first part, no constraints on
the end-to-end distance $R$ are assumed,
whereas the second part investigates the
effect of an externally imposed end-to-end distance $R$.

\subsection*{Unconstrained end-to-end distance}

There are three main consequences if the polymer (or part of it) is found in a
helical state: as the strand winds tightly around a central axis, the effective
polymer length is shortened dramatically by a factor $\gamma$, which is about
0.4 for a typical polypeptide.
Secondly, the hydrogen bonds prevent the monomers from free rotation, hence
increase the effective persistence length from about $a_c=1.8{\ \rm nm}$ in the
coil state to $a_h=200\ {\rm nm}$ in the helical state
\cite{poland-scheraga70,grosberg-khokhlov94}. Thirdly, as the number of
available configurations has decreased, the gain $\Delta h$ in potential energy
per monomer by forming hydrogen bonds competes with the associated loss of
entropy $\Delta s$. The net balance is established by writing down the free
energy per monomer:
\begin{equation}\label{monomer-energy}
    \Delta f=\Delta h - T \Delta s,
\end{equation}
where $T$ is the temperature. Since $\Delta h$ and $\Delta s$ are both
negative on formation of hydrogen bonds, we observe that at low temperature
the formation of helices is promoted, whereas at high temperatures the coil
state is favoured. Later, we shall have to distinguish two cases according to
the sign of the net free energy difference $\Delta f$.

At this stage, the model is far from being
exhaustive: monomers located at the ends of
helical domains suffer a great entropy reduction, but do not form hydrogen
bonds.
Therefore each interfacial monomer between a coil and helical domain
has an increased free energy of
\begin{equation}\label{interface-energy}
    \Delta f_t =-\Delta h,
\end{equation}
compared to a monomer in the helical state.
This additional interfacial energy contribution will
suppress domain boundaries, and hence render the helix-coil transition in this
model more cooperative.

In the literature, one normally gives the above model parameters $\Delta f$ and
$\Delta f_t$ in terms of the
Zimm-Bragg parameters $s$ and $\sigma$ \cite{zimm-bragg59}, which are merely the
exponentials of these quantities:
\begin{eqnarray}
    s      & = & \exp(- \beta\Delta f)\nonumber\\
    \sigma & = & \exp(-2\beta\Delta f_t),\label{zimm-bragg-par}
\end{eqnarray}
where $\beta=1/(k_{\rm B} T)$ with $k_{\rm B}$ the Boltzmann constant.
The factor 2 in eq. (\ref{zimm-bragg-par}) takes into account the fact that a
helical domain has two interfaces with the coil domains, hence two monomers
suffer from loss of configurational states, while they do not gain any
additional energy from hydrogen bonds.

After identifying the microscopic states of the chain monomers
$\{s_1 s_2 \dots s_N\}$, with
$s_i=c$ or $h$ for the mo\-no\-mer $i$ to be in a coil or helical state,
respectively, and their corresponding energetic contributions (see previous
paragraph), relevant quantities of interest, e.g. the average helical fraction
of an ensemble of polymers $\langle\chi\rangle$,
can be determined by the usual
statistical-mechanical calculation involving the  partition function
\begin{equation}
    Z=\sum_{\{s_1 s_2 \dots s_N\}} e^{-\beta F(\{s_1 s_2 \dots s_N\})},
        \label{part-sum}
\end{equation}
where the sum is performed over all possible states of the variables $s_i$,
$i=1,\dots,N$.
The total free energy $F$, depending on the states $\{s_i\}$, is the sum
of monomer free energy contributions (\ref{monomer-energy}) and
(\ref{interface-energy}). This
procedure will also take a mixing entropy into account: for a fixed number
of coil
and helical domains, the sequence can be changed without energy change, provided
that a helical domain follows a coil domain and vice versa. Therefore,
configurations of shorter domain structure receive a higher statistical weight.

Experimentally, the average helical fraction in a solution of free polymers is
accessible by measuring the net optical rotation of polarised light. However,
experiments \cite{teramoto01,yue-berry96,doty-bradbury56} deal with polymers in
solution in different environmental conditions: a change in temperature or in
another variable (pH value, salt concentration) affects the free energies
$\Delta f$ and $\Delta f_t$, hence by means of averaging over states, the
average helical fraction $\chi$ is altered. Qualitatively, one can observe that
at low temperature the optical rotation increases. Hence, one can conclude that
the segments condense into helical domains, whilst at high temperature thermal
fluctuations destroy the order and drive the segments preferably into the coil
state.

Buhot and Halperin \cite{buhot-halperin02} pursue the evaluation of
the partition
function (\ref{part-sum}) with great care and demonstrate that a me\-thod
merely based on a coarse-grained free energy approach leads to the same
result as the transfer
matrix me\-thod, the original treatment used by Zimm and Bragg
\cite{zimm-bragg59}, also used for example in
\cite{grosberg-khokhlov94,tamashiro-pincus01}.

\begin{figure}
\center
\resizebox{0.6\textwidth}{!}{\includegraphics{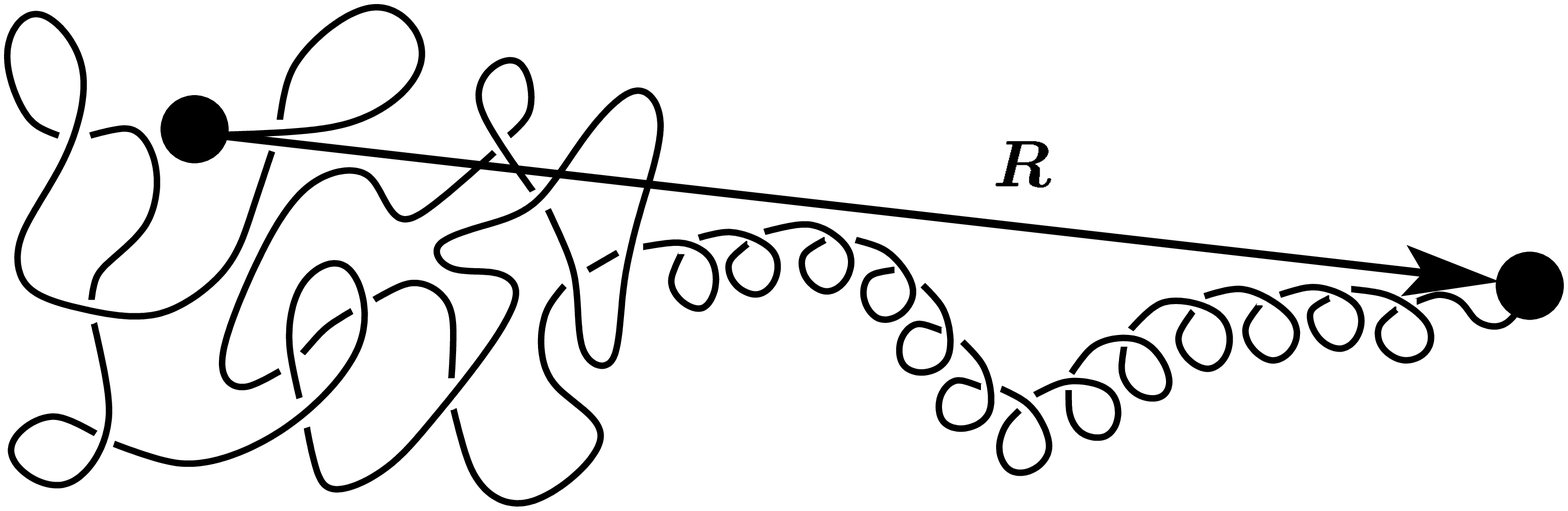}}
\caption{An illustration of the diblock approximation: neglecting the entropy
arising from the reordering of short helical segments is equivalent to assuming
that helical and coil domains are separated in two coherent domains.}
\label{fig2}
\end{figure}

In a first approximation, however, one can neglect the mixing entropy and
follow a very simple route. We assume in the following that the helical and coil
domains are artificially separated in just two blocks (cf. fig. \ref{fig2}).
This is equivalent to neglecting any
form of cooperativity in the system: firstly,
the entropy arising from reordering the domains (maintaining the alternating
succession of coil and helical domains) is neglected, and
secondly the interfacial
energy is similarly disregarded.

The detailed investigation by Buhot and Halperin
\cite{buhot-halperin02} showed that this
simplification does not change the outcome qualitatively and only mildly
quantitatively. Encouraged by their results, we apply a similar strategy
here and treat the strands in a diblock approximation. Since we now neglect the
effect of reordering the domains on the statistical weight, the system is
microscopically described by the fraction of helical segments $\chi$
($0\leq\chi\leq 1$).

\subsection*{Fixed end-to-end distance}

So far, the end-to-end distance $R$ has been unconstrained. If, however,
the chain has to span an externally imposed distance $R$ between its ends,
the system can only access a subset of configurations, in other
words, it will suffer a reduction of entropy. This argument applies
both to the coil and
helical part of the chain. However, the cost of reducing the entropy
is much higher for the coil part than for the helical part since most of the
configurational entropy in the helical part has been reduced anyway by the tight
interlocking (bonding) between adjacent segments. Therefore, the helical
segments become aligned between the externally fixed endpoints and the coil
segments are tautened over the distance which cannot be bridged by the helical
segment.

The free energy of the whole chain, under these assumptions,
reads as follows:
\begin{equation}\label{free-theta}
    F_{\rm ch}=\chi N\Delta f+ 2\Delta f_t + F_{\rm coil},
\end{equation}
where $N$ is the total number of monomers of length $a$. The free energy term
$F_{\rm coil}$ describes a coil polymer with end-to-end
distance $R$ minus the
amount of distance which can be bridged by the helical part.

To describe this last term $F_{\rm coil}$, we use the Gaussian approximation
of an ideal random walk. This approximation breaks down at high
extensions, but the practical quantitative deficiencies are not great as the
investigations in ref. \cite{buhot-halperin02} have revealed. We therefore have
\begin{equation}\label{coil-theta}
\beta F_{\rm coil} =
    \frac{3}{2}\frac{1}{(1-\chi)Na^2}\left(R-\gamma aN\chi\right)^2,
\end{equation}
taking into account the reduction in the contour length available for the coil
and also the reduction of the effective coil end-to-end distance.

Before moving on and introducing the basic techniques for a network theory, we
reproduce the findings for this model. The minimisation condition
$\partial F_{\rm ch}/\partial\chi=0$ allows us to determine the
equilibrium value $\chi=\chi(R)$. Subsequently, by substituting
$\chi(R)$ back into eqs. (\ref{free-theta}) and (\ref{coil-theta}),
we can obtain the effective free energy
$F_{\rm ch}(R)$. At this point, we have to
distinguish several cases.

We first look at the case $\Delta f>0$. In the relaxed state at $R=0$,
there is no spontaneous formation of helical segments,
simply because it would require a
positive free energy $\Delta f>0$ per segment to create a helix.

\begin{figure}
\center
\resizebox{0.45\textwidth}{!}{\includegraphics{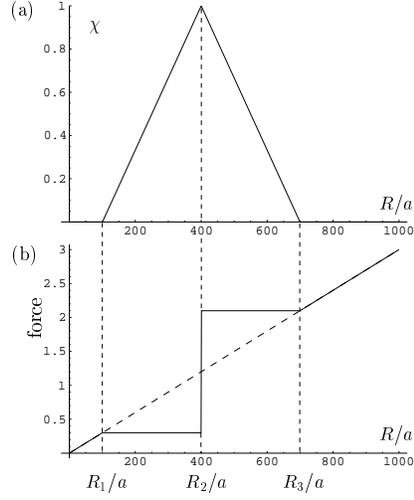}}
\caption{This pair of diagrams shows the helical fraction $\chi$ (a)
and the force (b), in units of $k_{\rm B} T/a$, vs.
the end-to-end distance $R$. The parameters
are as follows: number of monomers $N$=1000,
free energy of a helical state
$\beta\Delta f$=0.1, factor of shortening $\gamma$=0.4. The dashed line shows
the purely Gaussian behaviour of a free random walk. According to eqs.
(\ref{regime-boundaries}), $R_1=94.5a$,
$R_2=400a$ and $R_3=705.5a$.}
\label{fig3}
\end{figure}

By pulling the end points apart, i.e. by increasing $R$, one reduces the
entropy of the coil, hence raises the free energy of the whole system. At a
certain
extension $R_1$, the increase in the free energy due to stretching matches
the free
energy required to transfer a monomer into a helical state. If the extension $R$
is increased beyond $R_1$, then the molecule starts to form helical domains, at
a constant force (see fig. \ref{fig3}).

However, if the penalty for forming helical segments
$\Delta f$ is high, above a certain
critical value $\Delta f_{\rm crit}$ (to be given below), this effect does
not occur and the molecule stays in a coil state permanently, irrespective of
the imposed stretching.

We are left to consider the case $\Delta f<0$. At the statistically optimal
$R=0$, the
molecule goes spontaneously into the helical state, driven by the net free
energy gain. The end-to-end distance $R$ can be increased without a big energy
cost by aligning the helical segments. However, at an extension of
$R_2=\gamma a N$, the tight helical segments are torn apart and some
stretched coil domains are created.

\begin{figure}
\center
\resizebox{0.45\textwidth}{!}{\includegraphics{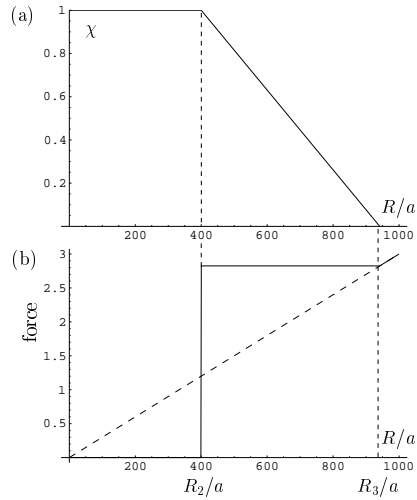}}
\caption{The analogous plot as in fig. \ref{fig3}, but with an attractive
interaction between helical monomers, meaning a negative free energy:
$\beta\Delta f=-0.2$. All other parameters
are the same as in fig. \ref{fig3}. Again, the dashed line indicates
simple Gaussian behaviour of a model random walk.}
\label{fig4}
\end{figure}

It is best to summarise the results in diagrams (Figs. \ref{fig3} and
\ref{fig4}), supplemented by analytical expressions. We show in two pairs of
two diagrams the helical fraction $\chi$ vs. end-to-end distance $R$ and the
force extension curve, obtained from $\partial F_{\rm ch}/\partial
R$.

It is useful to define the following auxiliary quantities:
$$
    \Delta R= a \gamma N \sqrt{1-\frac{\Delta f}{\Delta f_{\rm crit}}}
    \qquad
    \Delta f_{\rm crit}=\frac{3}{2}\gamma^2 k_{\rm B} T
$$

We then can determine the three regime boundaries in the extension coordinate
$R$:

\begin{equation}\label{regime-boundaries}
    R_1=R_2-\Delta R \quad
    R_2=\gamma a N \quad
    R_3=R_2+\Delta R
\end{equation}

The function $\chi=\chi(R)$ is defined in the obvious way (see Figs.
\ref{fig3} and \ref{fig4}). For the force
in response to the imposed extension $R$, one
obtains the following expressions
for the two cases, i.e. for the repulsive ($\Delta f>0$) and attractive
($\Delta f<0$) case respectively:
\begin{eqnarray*}\label{single-force}
{\rm force}|_{\Delta f>0}&=&k_{\rm B}T
    \begin{cases}
        \frac{3}{Na^2} R       & \text{for $0   \leq R < R_1$}\\
        \frac{3}{Na^2} R_1  & \text{for $R_1 \leq R < R_2$}\\
        \frac{3}{Na^2} R_3  & \text{for $R_2 \leq R < R_3$}\\
        \frac{3}{Na^2} R       & \text{for $R   \geq R_3$}
    \end{cases}
\\
{\rm force}|_{\Delta f<0}&=&k_{\rm B}T
    \begin{cases}
            0                  &  \text{for $0   \leq R < R_2$}\\
            \frac{3}{Na^2} R_3 &  \text{for $R_2 \leq R < R_3$}\\
            \frac{3}{Na^2} R   &  \text{for $R \geq R_3$}
    \end{cases}
\end{eqnarray*}

From these equations for the force, one obtains expressions for the
free energy $F(R)$ by integrating over $R$. These expressions
involve, besides some constants, linear and qua\-dra\-tic terms in $R$.

The model is based on the Gaussian approximation, used to describe the random
walk part of the chain. It assumes that the end-to-end distance of the coil
part is much
shorter than the coil's contour length. Therefore, the
approximation breaks down at high extensions: the true force diverges
\cite{treloar75}, however,
the force derived using the Gaussian model remains finite (see figs.
\ref{fig3} and \ref{fig4}). This artefact of the simplified
model will later be compensated by the fact
that, nevertheless, the Gaussian model correctly predicts a very low probability
of finding a chain at such large extensions. To illustrate this effect, we refer
to figs. \ref{fig3} and \ref{fig8}: the abrupt surge in the force at $R_2$
implies that the free energy $F(R)$ increases at a much higher rate for
$R\geq R_2$, hence the probability of finding a polymer with $R\geq R_2$
decays very rapidly.
A more realistic model with a divergent force and free energy would
only increase this effect.

\section{Network theory}\label{sec3}

The fact that the
polymer chains cannot intersect each other, im\-plies that topological
constraints and
entanglements  play a crucial role and alter the statistics significantly.
However, a very simple theory can reproduce many basic experimental
findings: the phantom chain network approach \cite{treloar75}.

This model assumes that the polymer strands actually do not interact along their
backbones, but only at their end points where they are linked to each other.
Hence, the strands are effectively transparent to each other and are
allowed to fluctuate
freely, with only the end-to-end distance $R$ constrained.
This assumption is not as naive as it might appear: there are deep physical
reasons why one can neglect self-interactions in a dense polymer system, and why
lateral (tube-like) constraints do not matter until a topological knot is tied
around the chain \cite{treloar75,doi-edwards86}.

Within the Gaussian model of a random walk,
the probability distribution of the end-to-end distance reads as usual
\cite{doi-edwards86}:
\begin{equation}\label{gauss-dis}
    P({\bm R})\propto \exp\left(-\frac{3}{2Na^2}{\bm R}^2\right),
\end{equation}
assuming that $|{\bm R}|\ll a N$, the chain contour length.

Accordingly, the free energy of a chain
is quadratic in $\bm R$ for small extensions:
\begin{equation}\label{gaussfree}
    \beta F=-{\rm ln} P({\bm R})=\frac{3}{2Na^2}{\bm R}^2,
\end{equation}
which im\-plies a simple Hookian behaviour in response to
stretching.

The distinctive feature of a network consists in memorising the initial
end-to-end distance distribution of chains as they are cross\-link\-ed
at their end-points.
In a simple network theory, one assumes that the polymers obey the equilibrium
melt statistics when they are crosslinked. Hence, the statistical distribution
of cross\-link\-ed network spans
$\bm R$ obeys the Gaussian form (\ref{gauss-dis}).
Since it turns out that for the overwhelming number of
chains the condition $|{\bm R}|\ll a N$ is satisfied,
the  quadratic form for the free energy per chain (\ref{gaussfree})
is well justified and can be used in the network calculation.
This leads immediately to the average total free energy per chain in a random
network:
\begin{equation}\label{netw-energy}
    F=\int d{\bm R}\ F({\bm R})\ P({\bm R}).
\end{equation}

Any external manipulation only affects the term $F({\bm R})$
in eq. (\ref{netw-energy}), whereas the distribution $P({\bm R})$ is left
un\-chang\-ed, it is said to be topologically quenched. For example, an
affine macroscopic deformation $\mat{\lambda}$ of the whole body,
which can serve as a good representation of the microscopic affine deformation
of the network strands ${\bm R}$, only
enters in the free energy, but leaves the distribution $P({\bm R})$
un\-chang\-ed. The elastic energy therefore becomes:
\begin{eqnarray}
    F_{\rm elast}\big(\mat{\lambda}\big)&=&
    \int d{\bm R}\  F(\mat{\lambda}{\bm R})\
        P({\bm R}),\nonumber\\
   {\rm with}\
    P({\bm R}) & \equiv &
    \frac{e^{-\beta F({\bm R})}}{\int d{\bm R}\ e^{-\beta F({\bm R})}}.
                                                            \label{elast-energy}
\end{eqnarray}
In general, the free energy $F({\bm R})$ does not need to be quadratic
(\ref{gaussfree}), leading to non-Gaussian distribution $P({\bm R})$.

An analogous reasoning will lead to the average helical content
$\langle\chi\rangle$ of a network,
which gives the total fraction of mo\-no\-mers in a helical state.
Under the assumption
that the end-to-end distance distribution at crosslinking obeys the equilibrium
distribution $P({\bm R})$ and is conserved in the crosslinked network,
an imposed deformation $\mat{\lambda}$ leaves the distribution
$P({\bm R})$ unaffected, but does enter $\chi(|{\bm R}|)$. Thus, the
dependence of the helical content $\chi$ on the deformation $\mat{\lambda}$
reads:
\begin{equation}\label{elast-theta}
    \langle\chi\rangle(\mat{\lambda})=
    \int d{\bm R}\ \chi\big(\big|{\mat{\lambda}}{\bm R}\big|\big)\
        P({\bm R}).
\end{equation}

\section{Network of helix-forming polymers}\label{sec-netw}

We now combine the theory for single molecules
with the quenched averaging to calculate network properties.
We will mainly focus on the total helical content described by
eq. (\ref{elast-theta}). Since the helical fraction $\chi$ depends on
the end-to-end distance $\bm R$, we
can expect that in a network, the average helical content
$\langle\chi\rangle$ would depend on the deformation $\mat{\lambda}$, giving an
opto-mechanical coupling.

To evaluate the integral in eq. (\ref{elast-theta}), note that all the
functions, i.e. $F(R)$ and $\chi(R)$ uniquely depend on $R$, but are
only analytically defined piecewise (e.g. see eq. ({\ref{force-cases}))
between the
limits $R=0$, $R_1$, $R_2$, $R_3$ and $Na$ (see appendix \ref{A-reg}).
On the other hand, any
incompressible deformation $\mat{\lambda}$, which we assume here, acts
differently on different strands: some strands are expanded, whereas others are
compressed, depending on the direction of the end-to-end vector $\bm R$ of a
given strand.

We simplify the problem further by looking at uniaxial deformations along the
$z$-axis only:
\begin{equation}\label{lambda-deform}
\mat{\lambda}=\left(
    \begin{array}{ccc}
        1/\sqrt{\lambda}& 0 & 0\\
        0 & 1/\sqrt{\lambda}& 0 \\
        0 & 0 &\lambda
    \end{array}\right).
\end{equation}

The end-to-end distance of a strand which is initially aligned at an angle
$\vartheta$ with the
$z$-axis is therefore scaled by the factor $\eta$:
\begin{eqnarray}
    R'=\big|\mat{\lambda}{\bm R}\big| & = & \eta R \nonumber \\
    {\rm with} \
    \eta(\lambda,\vartheta) & = &
            \sqrt{\lambda^2 \cos^2\vartheta+\frac{1}{\lambda}\sin^2\vartheta}.
            \label{etadef}
\end{eqnarray}

As a result, the uniaxial deformation, as in eq. (\ref{lambda-deform}),
affects the
end-to-end distance $R$ differently according to the angle $\vartheta$ between
the initial end-to-end distance
$\bm R$ and the $z$-axis of the imposed deformation. For example,
if $\lambda>1$, this
is tantamount to an extension by a factor $\lambda$ along the axis of
deformation ($\vartheta=0$) and to compression by $1/\sqrt{\lambda}$
perpendicular to the axis ($\vartheta=\pi/2$), see fig. \ref{fig5}.
\begin{figure}
\center
\resizebox{0.45\textwidth}{!}{\includegraphics{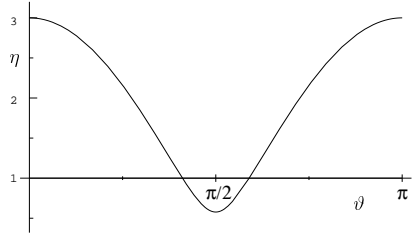}}
\caption{The elongation and compression $\eta$ of chains depends on the
polar angle $\vartheta$ (cf. eq. (\ref{etadef})). The scaling factor
$\eta$ becomes unity if $\arctan\vartheta=\sqrt{\lambda(\lambda+1)}$.
Here, we have chosen $\lambda=3$.}
\label{fig5}
\end{figure}

Clearly, one can see in fig. \ref{fig5} that at two polar angles, for which
$\tan\vartheta=\sqrt{\lambda(\lambda+1)}$, the end-to-end distance of the chains
is not affected by the deformation. In other words, the set of vectors
$\bm R$ invariant
under the deformation $\mat{\lambda}$ lies on a cone, see fig. \ref{fig6}.
\begin{figure}
\center
\resizebox{0.3\textwidth}{!}{\includegraphics{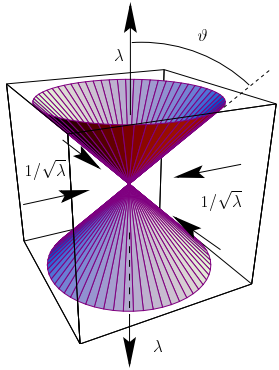}}
\caption{The locus of strands which are invariant under the uniaxial deformation
$\mat{\lambda}$ (\ref{lambda-deform}). The cone has an opening angle of
$\vartheta=\arctan\sqrt{\lambda(\lambda+1)}$. For $\lambda>0$, the arrows
indicate the approximate deformation of chains.}\label{fig6}
\end{figure}

The boundaries between which the function
$\chi\big(\big|\mat{\lambda}{\bm R}\big|\big)=\chi(\eta R)$ is analytically defined,
are similarly scaled
to $R_1/\eta$, $R_2/\eta$, or $R_3/\eta$, respectively.
For a specific value $\eta$, one
has to break up the radial $R$ integration in
\begin{equation}\label{avtheta}
    \langle\chi\rangle(\lambda) =
        \int d\Omega\!\!\!\!\!\!\!\!\!
        \int\limits_{[\rm piecewise]}\!\!\!\!\!\!\!\!\! dR\ R^2\ \chi(\eta R)\
        P(R)
        \equiv \int \!\! d\Omega\ I(\eta)
\end{equation}
not only
at the values $R=R_1$, $R_2$ and $R_3$, but also at the values $R=R_1/\eta$,
$R_2/\eta$ and $R_3/\eta$.

Depending on the value of $\eta$, the breakup values adopt a particular order
among themselves. At critical values
$$
    \eta_1=\frac{R_3}{R_2} \quad
    \eta_2=\frac{R_2}{R_1} \quad
    \eta_3=\frac{R_3}{R_1}
$$
some limits coincide, and as $\eta$ crosses these values, the order of breakup
limits is changed. We refer to the appendix \ref{A-reg} for an explicit
determination of regimes.

For any particular value of $\eta$, the $R$-integration can
be carried out giving analytical expressions in $\eta$
involving Error functions. But for varying $\eta$, the integration regimes
change, hence also the integrals. Therefore, the function $I(\eta)$ selects
among a set of functions $I^{\pm}_i (\eta)$, depending on the value of
$\eta$ (see appendix \ref{A-int}).
$$
    \langle\chi\rangle(\lambda)=\int d\varphi\int d\vartheta\
        \sin(\vartheta)\ I_{\rm select}(\eta(\lambda,\vartheta)),
$$
where $I_{\rm select}(\eta)$ selects the appropriate expression for
$I^{\pm}_i (\eta)$ depending on the value of $\eta$.

In the last step, the integration over the two angles $\varphi$ and
$\vartheta$ is carried out: the system is rotationally invariant, hence the
integration over the azimuthal angle $\varphi$ is trivial. The integration over
the polar angle $\vartheta$ is carried out numerically.

\section{Results and discussion}\label{sec-r-d}

We demonstrate the effect of network averaging
by going through a par\-ti\-cu\-lar example. We choose
the following parameters: $N$=1000 and $\gamma$=0.4,
and first examine the case of $\beta\Delta f$=0.1, that is when helix formation
is unfavourable in equilibrium. Note that, for convenience, we give all
quantities in
dimensionless units, expressing lengths in units of
the basic persistence length $a$.

On the one hand, the helical fraction $\chi$ of a single mo\-le\-cu\-le with
these parameters is peaked around
$R_2=400 a$ and the flanks end at $R_1=94.5 a$ and $R_3=705.5 a$ (see fig.
\ref{fig3}). On the other hand, the probability density distribution
\begin{equation}\label{prob-dens}
    P(R)=R^2 \exp(-\beta F(R)),
\end{equation}
which is valid for the single molecule statistics and also for
the network calculation, is peaked at a value $R^*=30.8 a$
(see fig. \ref{fig7}).
Therefore, the network average $\langle\chi\rangle$ for the undistorted
network ($\lambda=1$) will yield a very
low helical content, as there are only
very few chains present where $\chi(R)\not=0$.

\begin{figure}
\center
\resizebox{0.6\textwidth}{!}{\includegraphics{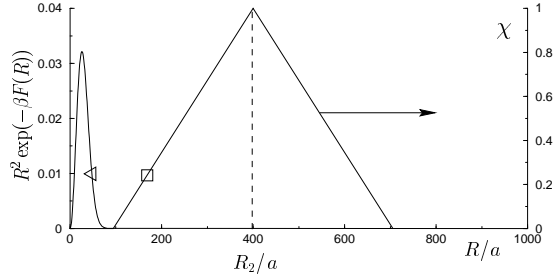}}
\caption{The probability density eq. (\ref{prob-dens}), labelled by a triangle
(left axis), and the helical fraction, labelled by a square (right axis).
The former is peaked at about $R=30.8 a$, the latter at $R=\gamma a
N=400$, leading to a very low helical content in the network (see dotted line
in fig. \ref{fig9}). The parameters
are those of fig. \ref{fig3}.}\label{fig7}
\end{figure}

However, if a uniaxial extension, parametrised by $\lambda$ (see eq.
(\ref{lambda-deform})) is applied, then some chains are
stretched and some are compressed (fig. \ref{fig6}). As we have seen
from eq. (\ref{prob-dens}), most of the chains are clustered near the shell of
radius $R^{*}$.
In pictorial terms, the uniaxial deformation now takes
chains which are both near this shell and close to the axis of
the deformation (i.e. around the poles of the $R^*$ shell) to a deformed
state: the chains will be clustered around two points on the axis at a distance
$\lambda R^*$ from the origin. As the parameter $\lambda$ becomes sufficiently
large, the uniaxial deformation  takes some increasing population of
chains into a region where the helical fraction $\chi(\mat{\lambda}{\bm R})$
becomes nonzero.

Taking this reasoning to an extreme, we can think of large values for $\lambda$,
where actually the chains are stretched to such a degree that they are
transferred beyond the peak of $\chi(R)$ (cf. fig. \ref{fig7}), that is,
$\lambda R^{*} \geq R_2 \equiv \gamma a N$.

Moreover, the end-to-end distance of chains with $\bm R$ perpendicular to the
axis is scaled by the factor $1/\sqrt{\lambda}$. Hence, for $\lambda>1$, these
chains are shrunken and thus, in our example, taken to a region where the
helical fraction $\chi(R)$ is still zero.

On the other hand, if we consider a deformation with $\lambda<1$, i.e. a
compression along the $z$-axis with a simultaneous elongation in the plane
perpendicular to the $z$-axis, then the roles are interchanged: for an
appropriate $\lambda$, chains with $\bm R$ approximately of length $R^*$
perpendicular to the axis are deformed to such a degree  that the helical
content $\chi(\mat{\lambda}{\bm R})$ becomes nonzero, whilst chains with
$\bm R$ approximately along the axis are taken to regions where
$\chi(\mat{\lambda}{\bm R})=0$.

Therefore, in the example considered (figs. \ref{fig3} and \ref{fig7}),
compression and elongation both lead to an increase in the helical content, the
latter however only at very high compression rates, hardly achievable in
practice.
The averaged helical content of this example is plotted in fig. \ref{fig9}
(dotted line).

Another choice of parameter values gives a completely different
behaviour. For an illustration of this,
we refer the reader to fig. \ref{fig8}, which shows the behaviour
of a chain with $\beta\Delta f=0.005$, a much lower free energy penalty for the
spontaneous helix formation. Now, the peak of the probability
distribution is around $R^*=159.2 a$, whereas the helical fraction reaches
the maximum at $R=400 a$. As these two function now show a much larger
overlap, the
average helical content in a network is much bigger, and furthermore, it can be
increased by both compression and elongation, see the dashed line in fig.
\ref{fig9}.

Note that the probability distribution in fig. \ref{fig8} justifies the Gaussian
approximation used earlier in eq. (\ref{coil-theta}):
for extensions above $R_2$,
the free energy has an increased slope, therefore, the probability distribution
shows a fast decay. In other words, chains at such high extensions correctly
contribute very little to the quenched averaging, eqs. (\ref{elast-energy})
and (\ref{elast-theta}).

\begin{figure}
\center
\resizebox{0.6\textwidth}{!}{\includegraphics{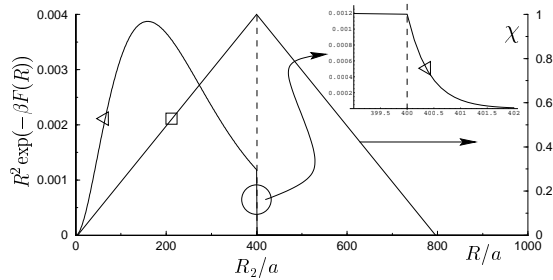}}
\caption{The analogous plot to fig. \ref{fig7}, but now for much smaller
$\beta\Delta f=0.005$. There is a sufficiently high overlap of the probability
distribution function and the helical fraction to generate a helical content in
the network (see dashed line in fig. \ref{fig9}). The inset shows the expanded
region of the transition at $R_2$.}\label{fig8}
\end{figure}

Fig. \ref{fig9} shows the different responses of the average helical
content in a
network for various positive values of the free energy $\beta\Delta f$.
For negative values of the free energy, that is for polymers spontaneously
folding into helices at given conditions,
fig. \ref{fig10} gives an analogous
set of results. Note that in the limit
$\beta\Delta f\rightarrow 0$, both cases converge
to a qualitatively similar curve, although our model does not treat this case
properly. In the limit of vanishing free energy penalty,
the two cases give a different
behaviour for the helical fraction of one single molecule. As
$\beta\Delta f \rightarrow 0^{+}$, the helical fraction $\chi$ looks similar
to \ref{fig3}a, where\-as for $\beta\Delta f \rightarrow 0^{-}$, $\chi$
resembles fig. \ref{fig4}a. This is the upshot of the fact that, in our model,
there is a discontinuous jump in the helical fraction as $\Delta f$ changes
sign.

Moreover, the results inferred for the case of a negative free energy
$\beta\Delta f$ have to be taken with some caution: the network theory,
outlined in section \ref{sec3}, assumes that the network consists of random
coils, allowing one to use Gaussian statistics.
However, for a negative free energy $\beta\Delta f$, the polymers
spontaneously form helical, semiflexible chains, which will obey different
statistics than flexible coil chains. Only for very long semiflexible chains,
the Gaussian statistics are again recovered.

\begin{figure}
\center
\resizebox{0.6\textwidth}{!}{\includegraphics{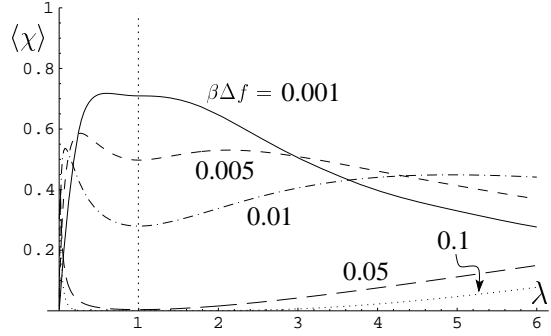}}
\caption{The helical content in a network for different parameters. The free
energy $\beta\Delta f$ takes the following values: 0.001 (solid line), 0.005
(dashed line), 0.01 (dot-dashed line), 0.05 (long-dashed line) and 0.1 (dotted
line); the other parameters ($N=1000$, $a=1$) are kept constant.}\label{fig9}
\end{figure}

\begin{figure}
\center
\resizebox{0.6\textwidth}{!}{\includegraphics{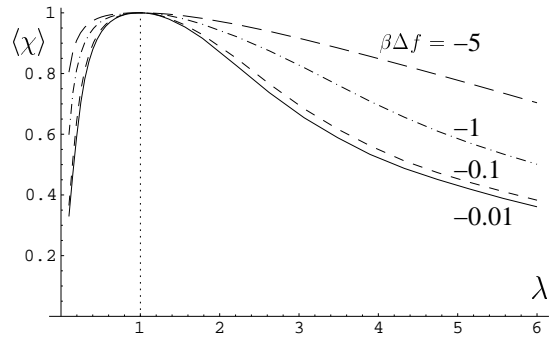}}
\caption{The helical content in a network for different values of the free
energy: $\beta\Delta f = -0.01$ (solid line), $-0.1$
(dashed line), $-1$ (dot-dashed line), and $-5$ (long-dashed line);
the other parameters ($N=1000$, $a=1$) are
kept constant.}\label{fig10}
\end{figure}

To summarise,
we have calculated the average helical content in a network based on
helicogenic polymer molecules described by a
simple coarse-grained model.
We found that a macroscopic deformation can have opposite effects on the helical
content of the network:
in the case of positive free energy $\beta\Delta f$, i.e.
the formation of helical
bonds leads to an overall increase of free energy (high temperature regime),
both an elongation and a compression can increase the overall helical content.
The extent of
this effect depends greatly on the parameters chosen for the polymer molecules.
In the case where the
formation of helical bonds is favoured by a free energy gain, the undeformed
network exhibits a maximal helical content, and any deformation, either
compression or elongation, decreases the helical content of the network.

We expect that these findings are relevant for investigations on biological
polymer networks by optical measurement, for example gelatin. In this material
however, the helices are formed by three strands which, on cooling, bind to each
other to act as effective crosslinks. Therefore, our model cannot be applied
straightforwardly. We have however demonstrated that microscopic effects of
macroscopic deformation can be observed in a network.

\section*{Acknowledgements}
S. K. enjoyed discussions with M. Rief and P. Nelson on this subject,
and would like to acknowledge support from Corpus Christi College,
Cambridge, from the Cavendish Laboratory, and from the Cambridge
Overseas Trust.

\appendix

\section{Regimes}\label{A-reg}

The integration domains of the $R$-integral in eq. (\ref{avtheta})
are broken up in several regimes. Here we show how these regimes are
determined and
the integration carried out. We only investigate the $R$-integration, although
we keep in mind that the factor $\eta$ can assume a range of values, namely
between $\lambda$ and $1/\lambda$ (cf. eq. (\ref{etadef}):
$$
    \langle\chi\rangle_R(\eta)=\!\!\!\!\!\!\!
    \int\limits_{[{\rm piecewise}]}\!\!\!\!\!\!\! dR\ R^2\ \chi(\eta R)\
    e^{-\beta F(R)}
$$
Note that the dependence on $\lambda$ and $\vartheta$ is hidden inside in the
variable $\eta$.

The comprehensive enumeration of all possible integrals is very lengthy, thus we
illustrate the general process by focusing on a particular example: we choose
the repulsive case ($\beta\Delta f>0$) in the second $\eta$-regime
with $\eta_1<\eta<\eta_2$ (see regime {\sc ii} in fig. \ref{fig3}).

The functions $\chi(\eta R)$ and $e^{-\beta F(R)}$ both can be expressed
analytically only in piecewise regimes:
\begin{eqnarray}
\chi(\eta R)&=&
    \begin{cases}
            0                                   & \text{for $0 \leq R < R_1/\eta$}\\
            1+\frac{1}{\Delta R}(\eta R-R_2)&
                                    \text{for $R_1/\eta \leq R < R_2/\eta$} \\
        1-\frac{1}{\Delta R}(\eta R-R_2)&
                                    \text{for $R_2/\eta \leq R < R_3/\eta$} \\
            0                                   & \text{for $R \geq R_3/\eta$}
    \end{cases}\label{force-cases}\\
\beta F(R)&=&
    \begin{cases}
        \frac{3}{2Na^2} R^2    & \text{for $0 \leq R < R_1$} \\
        \frac{3}{2Na^2} R_1^2+\frac{3R_1}{Na^2}(R-R_1)
                                        & \text{for $R_1 \leq R < R_2$} \\
        \frac{3}{2Na^2} R_3^2+\frac{3R_3}{Na^2}(R-R_3)
                                        & \text{for $R_2 \leq R < R_3$} \\
        \frac{3}{2Na^2} R^2    & \text{for $R \geq R_3$}
    \end{cases}\nonumber\\
    & & \label{def-eta-bF}
\end{eqnarray}

\section{Integrals}\label{A-int}

The restrictions on the right hand side of the eqs.
(\ref{def-eta-bF}) give the integration limits of the piecewise
integrations. For our particular choice of $\eta$
($\eta_1<\eta<\eta_2$), the integration path can be most easily
determined by looking at fig. \ref{figA1}, where a typical
integration path is highlighted together with the limits.
\begin{figure}[h]
\center
\resizebox{0.6\textwidth}{!}{\includegraphics{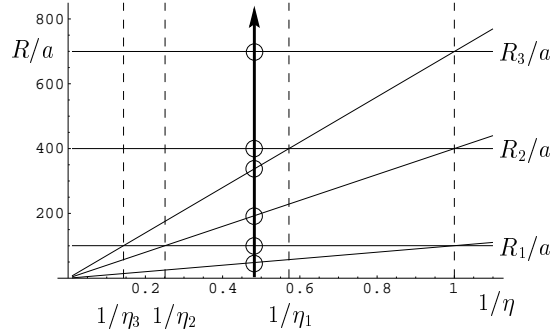}}
\caption{The integration regimes, parameters as in figure \ref{fig3}:
number of monomers $N$=1000 of length $a$=1,
net free energy difference for a monomer to be in a helical state
$\beta\Delta f$=0.1, factor of shortening $\gamma$=0.4. These parameters imply:
$1/\eta_1$=0.57, $1/\eta_2$=0.25 and $1/\eta_3$=0.14.}
\label{figA1}
\end{figure}

In this way, we are able to determine the appropriate limits of the integrals,
e.g:
\begin{eqnarray}
    I_{21}^{+}(\eta) & = &
        \int_{0}^{\frac{R_1}{\eta}} dR\ R^2\ \chi(\eta R)\ \exp(-\beta F(R))
                                                                                        \nonumber\\
    I_{22}^{+}(\eta) & = &
        \int_{\frac{R_1}{\eta}}^{R_1} dR\ R^2\ \chi(\eta R)\ \exp(-\beta F(R))
                                                                                        \nonumber\\
    & \vdots &\nonumber \\
    I_{27}^{+}(\eta) & = &
        \int_{R_3}^{\infty} dR\ R^2\ \chi(\eta R)\ \exp(-\beta F(R))
                                                        \label{piecewise-int}
\end{eqnarray}

We then put the piecewise integrations together again to obtain:
\begin{equation}\label{sumint}
    I_{2}^{+}(\eta)=\sum_{i=1}^{7}I_{2i}^{+}(\eta).
\end{equation}

Note that in the case of a net repulsive interaction ($\beta\Delta f<0$),
$\chi(\eta R)$ is zero for values $R<R_1$ and $R>R_3$. Therefore, the
integrals in eqs. (\ref{piecewise-int}) vanish in these regimes, and the
sum in (\ref{sumint}) reduces to three terms only.

For a different values of $\eta>1$, the procedure is exactly the same, giving
$I_{1}^{+}(\eta)$, $I_{2}^{+}(\eta)$, etc. and analogously for
values $\eta<1$, we obtain $I_{2}^{-}(\eta)$, etc.


\begin{thebibliography}{10}

\bibitem{smith-finzi92}
S. B. Smith, L. Finzi and C. Bustamante, Science {\bf 258}, 1122 (1992).

\bibitem{rief-oesterhelt97}
M. Rief {\em et al.}, Science {\bf 275},  1295  (1997).

\bibitem{oesterhelt-rief99}
F. Oesterhelt, M. Rief, and H.~E. Gaub, New J. Phys. {\bf 1},  6.1  (1999).

\bibitem{rief-grubmuller02}
M. Rief and H. Grubmuller, Chemphyschem {\bf 3},  255  (2002).

\bibitem{zimm-bragg59}
B.~H. Zimm and J.~K. Bragg, J. Chem. Phys. {\bf 11},  526  (1959).

\bibitem{poland-scheraga70}
D. Poland and H.~A. Scheraga, {\em Theory of helix-coil transitions in
  biopolymers} (Academic Press, New York, 1970).

\bibitem{saroff-kiefer99}
H.~A. Saroff and J.~E. Kiefer, Biopolymers {\bf 49},  425  (1999).

\bibitem{bloomfield99}
V.~A. Bloomfield, Am. J. Phys. {\bf 67},  1212  (1999).

\bibitem{teramoto01}
A. Teramoto, Prog. Polym. Sci. {\bf 26},  667  (2001).

\bibitem{yue-berry96}
S. Yue, G.~C. Berry, and M.~M. Green, Macromolecules {\bf 29},  6175  (1996).

\bibitem{doty-bradbury56}
P. Doty, J.~A. Bradbury, and A.~M. Holtzer, J. Am. Chem. Soc. {\bf 78},  947
  (1956).

\bibitem{grosberg-khokhlov94}
A.~Y. Grosberg and A.~R. Khokhlov, {\em Statistical physics of macromolecules}
  (AIP Press, New York, 1994).

\bibitem{buhot-halperin02}
A. Buhot and A. Halperin, Macromolecules {\bf 35},  3238  (2002).

\bibitem{buhot-halperin00}
A. Buhot and A. Halperin, Phys. Rev. Lett. {\bf 84},  2160  (2000).

\bibitem{tamashiro-pincus01}
M.~N. Tamashiro and P. Pincus, Phys. Rev. E {\bf 63},  021909  (2001).

\bibitem{treloar75}
L.~R.~G. Treloar, {\em The physics of rubber elasticity} (Clarendon Press,
  Oxford, 1975).

\bibitem{doi-edwards86}
M. Doi and S.~F. Edwards, {\em Theory of Polymer Dynamics} (Clarendon Press,
  Oxford, 1986).

\end{thebibliography}
\end{document}